\documentclass[fleqn,usenatbib]{mnras}

\usepackage{graphicx}
\usepackage{times}
\usepackage{amssymb} 
\usepackage{txfonts}
\usepackage{appendix}
\usepackage{multirow}
\usepackage{natbib}
\usepackage{longtable, portland}
\usepackage{supertabular}
\usepackage{rotating}
\usepackage{array}
\usepackage{caption}
\usepackage{gensymb}
\usepackage[T1]{fontenc}
\usepackage{color}

\title[Green valley active and inactive galaxies]{Star formation of far-IR AGN and non-AGN galaxies in the green valley: possible implication of AGN positive feedback}

\author[Antoine Mahoro]{Antoine Mahoro $^{1, 2}$\thanks{E-mail:
mahoroan@gmail.com}, Mirjana Povi\'c$^{3, 4}$, and Pheneas Nkundabakura$^{1}$\\
$^{1}$University of Rwanda, College of Education, School of Education, MSPE Department, P.O. Box 5039, Kigali, Rwanda\\
$^{2}$Mbarara University of Science and Technology, Faculty of Science, Physics Department, P.O. BOX: 1410 Mbarara, Uganda\\
$^{3}$Ethiopian Space Science and Technology Institute (ESSTI), Entoto Observatory and Research Center (EORC), \\
Astronomy and Astrophysics Research Division, P.O. Box 33679, Addis Ababa, Ethiopia\\
$^{4}$Instituto de Astrof\'isica de Andaluc\'ia (IAA-CSIC), Glorieta de la Astronom\'ia s/n, 18008 Granada, Spain}

\date{Accepted XXX. Received YYY; in original form ZZZ}

\pubyear{2017}

\begin{document}
\label{firstpage}
\pagerange{\pageref{firstpage}--\pageref{lastpage}}
\maketitle

\begin{abstract}
In this paper we present the star formation properties of $\rm{I_{subaru}\leq}$ 23 X-ray detected AGN and non-AGN galaxies in the green valley with far-IR (FIR) emission, using data from the COSMOS field. We measured star formation rates (SFR) using FIR Herschel/PACS data and we observed the location of AGN and non-AGN galaxies on the main-sequence of star formation. We went a step further in analysing the importance of AGN in quenching star formation in the green valley, the region proposed to be the transitional phase in galaxy evolution where galaxies are moving from later- to earlier-types. We found that most of our green valley X-ray detected AGN with far-IR emission have SFRs higher than the ones of inactive galaxies at fixed stellar mass ranges, the result that is different when considering optical data. These FIR AGN have still very active star formation, being located either on or above the main-sequence of star formation (in total 82\% of our sample). Therefore, they do not show signs of star formation quenching, but rather its enhancement. Our results may suggest that for X-ray detected AGN with FIR emission if there is an influence of AGN feedback on the star formation in the green valley the scenario of AGN positive feedback seem to take place, rather than the negative one.
\end{abstract}




\section{Introduction}
The bi-modality in the galaxy properties (e.g., colours, absolute magnitude, stellar mass, star formation (hereafter SF), etc.) was studied at different redshifts and wavelengths \citep[e.g.][]{kauffmann03, baldry04, salim07, brammer09, povic13, walker13, Lee}, showing the difference between the two main morphological groups: late-type and early-type galaxies (or in terms of star formation often called star-forming and quiescent galaxies, respectively). In commonly used colour-magnitude (or colour-stellar mass) diagrams, the regions where the majority of these two types are located are called ``blue cloud'' and ``red sequence'', respectively. The term ``green valley'' was introduced to represent the region between these two \citep{martin05}, and represents the transitional phase in galaxy evolution. Its study is therefore crucial for understanding the process of SF quenching and how galaxies transform from late- to early-types.    

This transition must be on more rapid timescales than those typical of blue cloud and red sequence, since a lower density of sources is present at all studied redshifts and wavelengths. However, different works reported different results. Some found a short timescale of less than 1Gyr \citep{faber07, Pan}, while others suggested non-single transitional states and reported from intermediate timescales (1\,-2\,Gyr) to slow-quenching \citep{Schawinski, Smethurst}.

Different mechanisms were studied and proposed as responsible for SF quenching and galaxy transformation, such are: active galactic nuclei (AGN) negative feedback \citep[e.g.][]{dimatteo05, Nandra}, environmental effects (e.g., ram pressure striping, thermal evaporation, starvation, strangulation, etc.) and cluster interactions \citep{gunn72, cowie77, larson80, bekki02,fang13}, galaxy minor and major mergers \citep[][and references therein]{barro13, Smethurst}, supernovae winds \citep[][and references therein]{marasco12}, and/or secular evolution \citep{mendez11, Smethurst}. 

How important is the role of AGN in SF quenching and what is the possible connection between them is still an open question. Different works found that most of X-ray detected AGN lie in the green valley, suggesting that the AGN feedback mechanism may play an important role in quenching SF and moving galaxies from the blue cloud to red sequence \citep{Sanchez2004, Nandra, Georgakakis2008, Silverman2008, Treister, Povi2, Povi}. Other results however found that there is no strong evidence in AGN host galaxies of either highly suppressed or elevated SF when compared to galaxies of similar stellar masses and redshifts (Alonso-Herrero et al. 2008; Brusa et al. 2009). A study of X-ray selected AGN in the green valley region revealed that AGN are being hosted with both early- and late-type galaxies, but with similar colours \citep{Povi2, Povi}. 

In this work we went a step further in analysing the importance of AGN in SF quenching. Using a sample of green valley non-active (hereafter non-AGN) and X-ray detected active galaxies, we analysed their star formation rates (SFR), measured by using far-infrared (FIR) data, and their location on the main sequence of star-forming galaxies. We used the data from the Cosmic Evolution Survey (COSMOS), with multiwavelength information available \citep{scoville07}. 

The paper is organised as follows: in Section 2, we explain the data used and the green valley sample selection. Spectral energy distribution (SED) fittings and SFRs and stellar mass measurements are explained in Section 3. Our main results are showed in Section 4, and discussed in Section 5. Throughout this paper we used a standard cosmology ($\Omega_{m}=0.3,\,\Omega_{\Lambda}=0.7$), with $H_{0}=70$\,km\,s$^{-1}$\,Mpc$ ^{-1}$. The stellar masses are given in units of solar masses (M$_{\odot}$), and both SFRs and stellar masses assume \citet{Salpeter1955} initial mass function (IMF).

\section{The Data}\label{Data}
The photometric data used in this work were taken from COSMOS \citep{scoville07} database, a 2 sq. deg. deep survey centered at RA\,(J2000)\,=\,10:00:28.6 and DEC\,(J2000)\,=\,+02:12:21.0. COSMOS was observed at all wavelengths, including the Advanced Camera for Surveys (ACS) on board of the Hubble Space Telescope (HST). More than 1.2 million sources were detected down to a limiting magnitude of I$\rm{_{AB}}$\,$\simeq$\,26.0 \citep{scoville07}.

For galaxy selection we used the morphological catalogue of \citet{Tasca2009}, which is based on the ACS photometric catalogue of \citet{Leauthaud}. The catalogue contains morphological classification of 237912 galaxies and is complete down to a magnitude I$\rm{_{AB}}$\,$\simeq$\,23.0. The morphological classification was done using three different non-parametric methods \citep{Abraham, Huertas, Tasca2009}, and all three classifications are available in the catalogue. The photometric redshifts (See Figure ~\ref{fig_GV_Phot}) were extracted from \citet{Ilbert2009} catalogue, and were computed using the LePhare code \citep{Arnouts2011} with 30 photometric bands down to a magnitude of I$\rm{_{AB}}$\,$\simeq$\,25.0. In the case of active galaxies (see below) we extracted the photometric redshifts from \citet{Salvato} catalogue, measured with the same code but using AGN templates.

Far-infrared (FIR) data used in this work come from observations done by the Photoconductor Array Camera and Spectrometer (PACS) instrument \citep{Poglitsch} on board of the Herschel Space Observatory, as part of the PACS Evolutional Probe (PEP, \citet{Lutz}) survey. As part of this survey the COSMOS field was observed at 100\,$\micron$ and 160\,$\micron$ over the full 2 sq. deg. field. The catalogue was extracted with 24 $\micron$ prior from the Spitzer data release 1 (DR1; \citet{Rieke}), and contains MIPS\footnote{Multi-band Imaging and Photometer for Spitzer} FIR and Herschel/PACS data. The detection limits of the COSMOS 24\,$\micron$ and Herschel catalogs are 80\,mJy, and 5 mJy and 10.2 mJy at 100\,$\micron$ and 160\,$\micron$ PACS bands down to 3$\sigma$, respectively (for further information regarding the PEP catalogue see \citet{Lutz}).

In X-ray the COSMOS field was observed by both Chandra and XMM-\textit{Newton}. In this work we used the Chandra catalogue of \cite{Civano2012} with 1761 X-ray sources with optical counterparts, observed within the central 0.9\,sq.deg. of the COSMOS field down to limiting flux of 7.3\,$\times$\,10$^{−16}$\,erg\,cm$^{−2}$\,s$^{−1}$ in the 2.0\,-\,10.0 keV band. The additional sources were extracted from \citet{Brusa2007} catalogue of 695 X-ray sources with optical counterparts detected in the first 1.3\,sq.deg. of the COSMOS XMM-Newton survey, down to limiting flux of 5\,$\times$\,10$^{−15}$\,erg\,cm$^{−2}$\,s$^{−1}$ in the 2.0\,-\,10.0 keV band.

\section{Sample selection}

\subsection{The AGN and non-AGN samples}
We used X-ray data to select AGN sample. We combined the two X-ray catalogues described above. Since the Chandra data are deeper, we kept all sources from \cite{Civano2012} and added the missing ones detected with XMM-Newton \citet{Brusa2007}. The obtained catalogue was than cross-matched with \citet{Salvato} photometric redshift catalogue. Finally, we used the flux ratio between the hard X-rays (2\,-\,10 keV) and optical I band: log[F$_{x}$/F$_{0}$]\,=\,log\,F$_{x}$\,+\,F$_{0}$/2.5\,+\,5.325, 
and selected as AGN all sources with $\rm{-1\leq\,log\,F_{x}/F_{0}\leq1}$ \citep{Alexander, Bauer, Bundy, Trump}. We obtained a final sample of 1,472 AGN.

For selecting non-AGN galaxies we used \cite{Tasca2009} catalogue. We first excluded all sources classified as stars in \citet{Leauthaud}. We then excluded all AGN, obtaining the final sample of 221,154 non-AGN sources.  

\subsection{The green valley galaxies}

\begin{figure}
\centering
\includegraphics[width=\columnwidth]{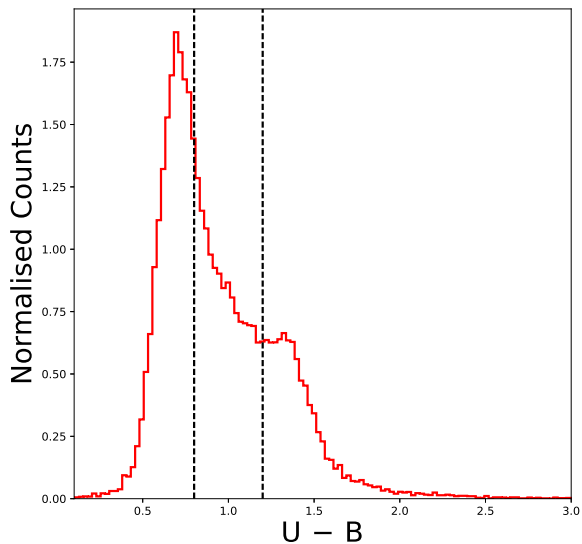}
\caption{Rest-frame U-B colour distribution of the total sample of COSMOS galaxies. Green valley is marked with black dashed vertical lines.}
\label{fig_GV}
\end{figure}

\begin{figure}
\centering
\includegraphics[width=\columnwidth]{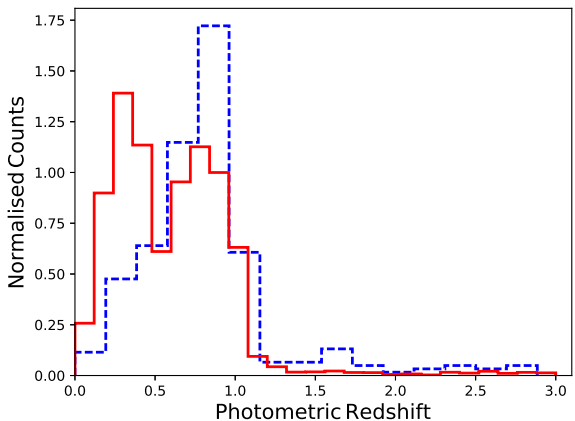}
\caption{Normalised distributions of photometric redshifts of green valley AGN (blue dashed line) and non-AGN (red solid line).}
\label{fig_GV_Phot}
\end{figure}

To select AGN and non-AGN galaxies located in the green valley we used the U\,-\,B rest-frame colours and criteria $\rm{0.8\leq U-B\leq 1.2}$. This criteria was already used in previous works for samples of galaxies at similar redshifts \citep[e.g.][]{Nandra, willmer07}. We initially run KCORRECT code \citep{Blanton} to apply the k-correction on both CFHT\footnote{Canada-France-Hawaii Telescope} $\rm{u}$ and Subaru $\rm{B}$ bands. Figure~\ref{fig_GV} shows the distribution of all galaxies and the limit that corresponds to our GV selection (black dashed vertical lines). As mentioned above the sample is complete down to magnitude $I\rm{_{AB} \leq 23}$. We selected in total 317 and 13,877 AGN and non-AGN green valley galaxies, respectively. This corresponds to 21.5\% and 6.3\% of total previously selected AGN and non-AGN galaxies, respectively. Figure~\ref{fig_GV_Phot} shows the photometric redshift distribution of all green valley galaxies. 88\% and 89\% of AGN and non-AGN, respectively, have redshifts between $0.2\,\leq\,z\,\leq\,1.2$. AGN sample shows higher values with median redshift of 0.8, while the median one for non-AGN FIR selected green valley galaxies is 0.54. By carrying out the Kolmogorov-Smirnov test we obtained that the two distributions are significantly different (having probability parameter of 2.49\,$\times\,10^{-7}$).

Taking into account the results from previous works where AGN were studied in the green valley, we do not expect significant AGN contamination on measured colours in our sample of active galaxies \citep[e.g][]{Kauffmann2007, Nandra, Silverman2008, Povi2}. To test the effect of dust extinction, we applied the same method as in \cite{FernandezLorenzo10} and measured B\,$-$\,V rest-frame colours before and after the extinction correction. We found that only $<$\,5\% of our galaxies might be removed from green valley region due to dust extinction. With this result we do not expect to have significant effect of dust extinction on our green valley sample selection.

\subsubsection{FIR green valley galaxies}

To measure the SFRs of AGN and non-AGN green valley galaxies (see the next section), we cross-matched the previously obtained samples with Herschel/PACS FIR data (see above). We first removed from PEP catalogue all sources with flag = -99\footnote{The flag ``-99'' has been used in the PEP catalogue in the columns of flux and flux error to indicate that there is no data, or that pixels are flagged as saturated in this band at this position. It was used as non-detection convention.} in both 100 and 160 $\rm{\micron}$ bands \citep{Laigle}. Finally, using the cross-matched radius of 2\,arcsec we obtained 114 and 2958 FIR green valley AGN and non-AGN sources, respectively. This means that of all green valley selected AGN and non-AGN sources, only 36\% and 21\% were detected with at least one Herschel/PACS band. 2\,arcsec cross-matching radius was selected as the best solution between avoiding bad matches and losing the possible counterparts, after carrying out the tests for different radii between 1 and 5\,arcsec. 

\begin{figure}
\centering
\includegraphics[width=\columnwidth]{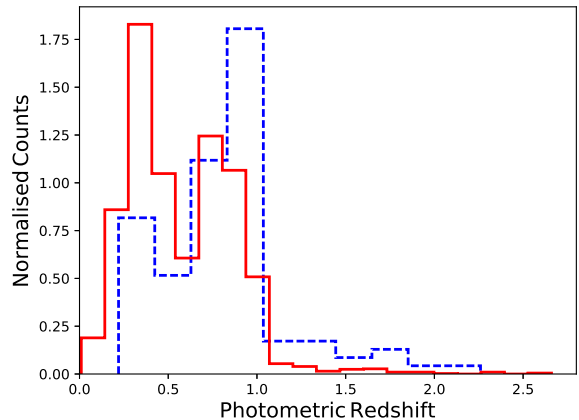}
\caption{Normalised distributions of photometric redshifts of FIR green valley AGN (blue dashed line) and non-AGN (red solid line) galaxies.}
\label{fig_GV_Phot_FIR}
\end{figure}

Figure~\ref{fig_GV_Phot_FIR} shows the photometric redshift distribution of selected FIR green valley galaxies. 89\% and 93\% of AGN and non-AGN, respectively, cover the range of $0.2\,\leq\,z\,\leq\,1.2$. Once again AGN sample shows higher redshifts, and the two distributions are significantly different (having the Kolmogorov-Smirnov probability parameter of 3.77 $\times\,10^{-25}$).

\section{SED fittings and SFR measurements}

SFRs were measured using Herschel/PACS FIR data. For all non-AGN, we assumed that all non-AGN the FIR luminosity is due to star formation, and that the total IR SF luminosity (L$\rm{_{IR}}$, the SF luminosity integrated over the range 8 - 1000 $\micron$) is dominated by the FIR luminosity \citep{povic2016}. The SFR is then derived from the total L$_{IR}$ according to the \citet{Kennicutt} relation modified for the Salpeter IMF \citep{Salpeter1955}:
\begin{equation}
\rm{\frac{SFR}{M_{\odot}yr^{-1}}\,=\,1.7\times 10^{-10}\bigg(\frac{L_{IR}}{L_{\odot}}\bigg)}
\label{SFR}
\end{equation}
and assuming $\rm{L_{\odot}\,=\,3.8\,\times\,10^{33}\,erg\,s^{-1}}$
To obtain the $\rm{L_{IR}}$ value of each source we carried out a spectral energy distribution (SED) fitting using the Le Phare\footnote{http://www.cfht.hawaii.edu/~arnouts/lephare.html} code \citep{Arnouts, Ilbert}, which separately fits the optical/NIR part of the spectrum with a stellar library and the MIR/FIR part (at $\rm{\lambda>7\micron}$) with IR libraries and gives the total L$_{IR}$. \\
\indent To fit the IR part of the SED, through $\chi^2$ minimisation, we used different templates for non-AGN and AGN samples. FIR green valley non-AGN galaxies were fitted using Chary \& Elbaz (2001) libraries. We checked visualy all fits and considered as the good ones only those where the best fit in MIR/FIR part passes through all available photometric points (including their errors). Finally, we obtained in total good fits for 2609 FIR green valley non-AGN (out of 2958). For FIR green valley AGN we used a set of 11 templates of \citet{Kirkpatrick2015}, with a known AGN contribution from 0\% up to 57.2\%. After checking visualy all fits, we obtained in total good fits for 103 FIR green valley AGN (out of 114). Of those, 37 sources show 1\% AGN contribution, 8 sources 4.2\%, 11 sources 6.7\%, 6 sources 9.1\%, 8 sources 15\%, 6 sources 21.4\%, 10 sources 25\%, 12 sources 39.1\%, and 5 sources 57.2\%. Table 3 in \citet{Kirkpatrick2015} lists the LIR for each template and the LIR(SF), which is the amount of LIR due to SF. The ratio $\rm{frac_{SF}\,=\,LIR(SF)/LIR}$ gives then the fraction of LIR due to SF for each template. We applied that fraction to the LIR of each source measured by Le Phare. Therefore, for an individual galaxy, LIR(SF) = LIR $\times$ frac$_{SF}$ (where frac$_{SF}$ comes from the best fitting template). Finally, we used the corrected LIR to measure SFRs of AGN (by using Equation ~\ref{SFR}).

\begin{figure}
\centering
\includegraphics[width=\columnwidth]{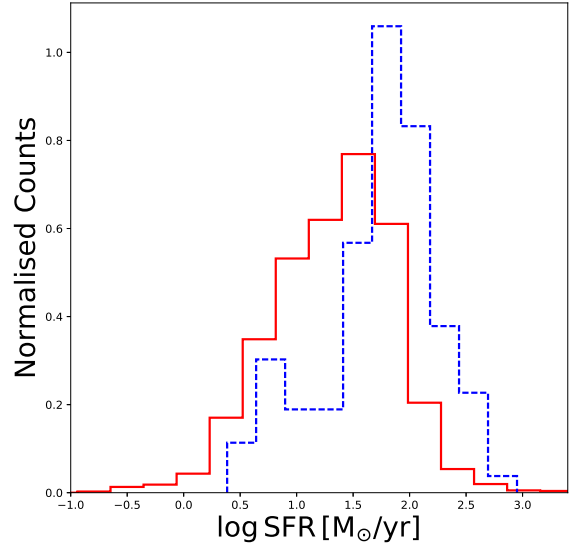}
\caption{Normalised distributions of SFR measurements of FIR green valley AGN (blue dashed histogram) and non-AGN (red solid histogram) samples.
\label{HIST}}
\end{figure}

Figure~\ref{HIST} shows the SFR distributions of FIR green valley AGN and non-AGN sources. We found that in general, when comparing whole samples, AGN show higher SFRs. 50\% of AGN cover the SFR range between 1.8 - 3.0\,M$_{\odot}$yr$^{-1}$, with a median of 2.11\,M$_{\odot}$yr$^{-1}$, while 50\% of non-AGN have SFRs between 1.35 - 2.6\,M$_{\odot}$yr$^{-1}$ and median SFR of 1.69\,M$_{\odot}$yr$^{-1}$ (for additional comparisons see Sec. 6).

We compared our SFRs with those measured by \citet{Whitaker}\footnote{\citet{Whitaker} measured SFR of 22,816 galaxies from the NOAO Extremely Wide-Field Infrared Imager Medium-Band (NEWFIRM) Survey, using the optical and NIR/MIR data up to 11\micron, and covering a part of the COSMOS and AEGIS fields.}. In the case of non-AGN, we found a strong correlation between the two measurements (with correlation coefficient of 0.917). In the case of AGN our SFRs are a bit lower than the ones obtained by \citet{Whitaker} for the same sample, with correlation coefficient of 0.765, which might be attributed to the fact that the \citet{Whitaker} measurements may suffer from the AGN contamination in NIR/MIR \citep{Donley,Rujopakarn}.

\section{Main sequence of star formation of FIR green valley AGN and non-AGN galaxies}

\begin{figure}
\centering
\includegraphics[width=\columnwidth]{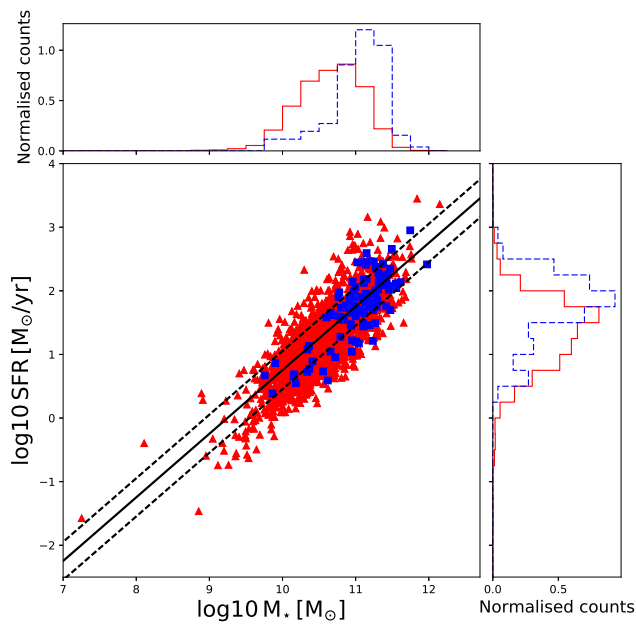}
\caption{Relation between the SFR and stellar mass for green valley galaxies. Non-AGN are represented by red triangles, while AGN are represented by blue squares. The solid black line shows the \citet{Elbaz2011} fit for the MS of star-forming galaxies observed with Herschel, while the dashed lines represent the typical MS width of $\pm$0.3 dex. Top and right histograms show the normalised distributions of stellar mass and SFR respectively, and comparison between the AGN (blue dash lines) and non-AGN (red solid lines) samples. \label{MS_both}}
\end{figure}

The common way to analyse the SF properties of galaxies is to study the relation between SFR and stellar mass, and to compare the location of sources with the main sequence (MS) of sfar-forming galaxies \citep[e.g][]{Brinchmann, Elbaz, Gonzalez2010, Whitaker, Guo2013, Leslie2016, povic2016, Netzer2016}. Figure~\ref{MS_both} shows all FIR green valley AGN and non-AGN sources studied in this work on the SFR-stellar mass diagram. For the SFRs we used FIR measurements from previous section. Stellar masses were measured through SED fitting, using the KCORRECT code \citep{Blanton}, the photometric information from 10 optical/NIR bands, $\sim$\,500 spectral templates \citep{Blanton}, and Salpeter IMF \citep{Salpeter1955}. For representing the MS of star-forming galaxies, we used the fit obtained by \citet[][and their equation 13]{Elbaz2011}, which was derived using a sample of galaxies observed with Herschel. The solid black line in Figure ~\ref{MS_both} shows this fit for z= 0.6, which is the average redshift in our sample. For the width of MS we used $\pm$ 0.3 dex (dashed line), found in many previous works to be the typical 1$\sigma$ value \citep[e.g,][]{Elbaz, Whitaker,Whitaker2014, Shimizu}. Table ~\ref{Table1} summarises the fractions of our FIR green valley AGN and non-AGN galaxies respect to the MS of SF, showing that $\sim$\,70\% of AGN and non-AGN are located on the MS. In addition, a slightly higher fraction of AGN was found above the MS than in the case of non-AGN galaxies.     

\begin{table}
\caption{Fraction of our green valley FIR AGN and non-AGN respect to the MS of SF from \citet{Elbaz2011}.}
\label{Table1}
\begin{tabular}{cccc} \hline
           & Below\,MS  & On\,MS$^*$  & Above\,MS \\\hline
  AGN      & 14\%         & 68\%  &  18\%      \\
  Non-AGN  & 9\%         & 70\%  &  21\%       \\ \hline
\end{tabular}
\begin{flushleft}
{* Within 1$\sigma$}
\end{flushleft}
\end{table}

\begin{figure*}
\centering
\begin{minipage}[c]{0.49\textwidth}
\includegraphics[width=7.5cm,angle=0]{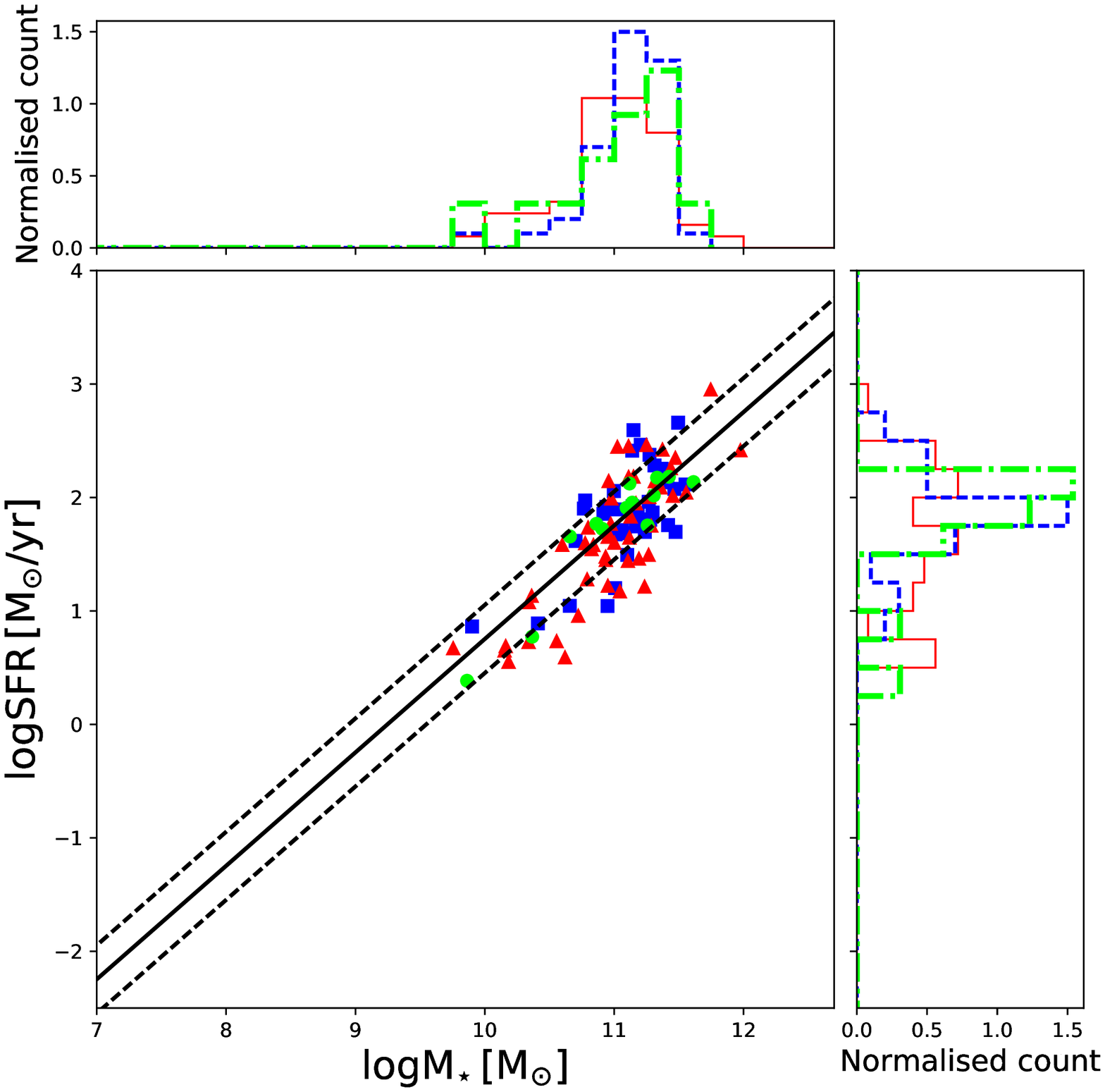}
\end{minipage}
\begin{minipage}[c]{0.49\textwidth}
\includegraphics[width=7.5cm,angle=0]{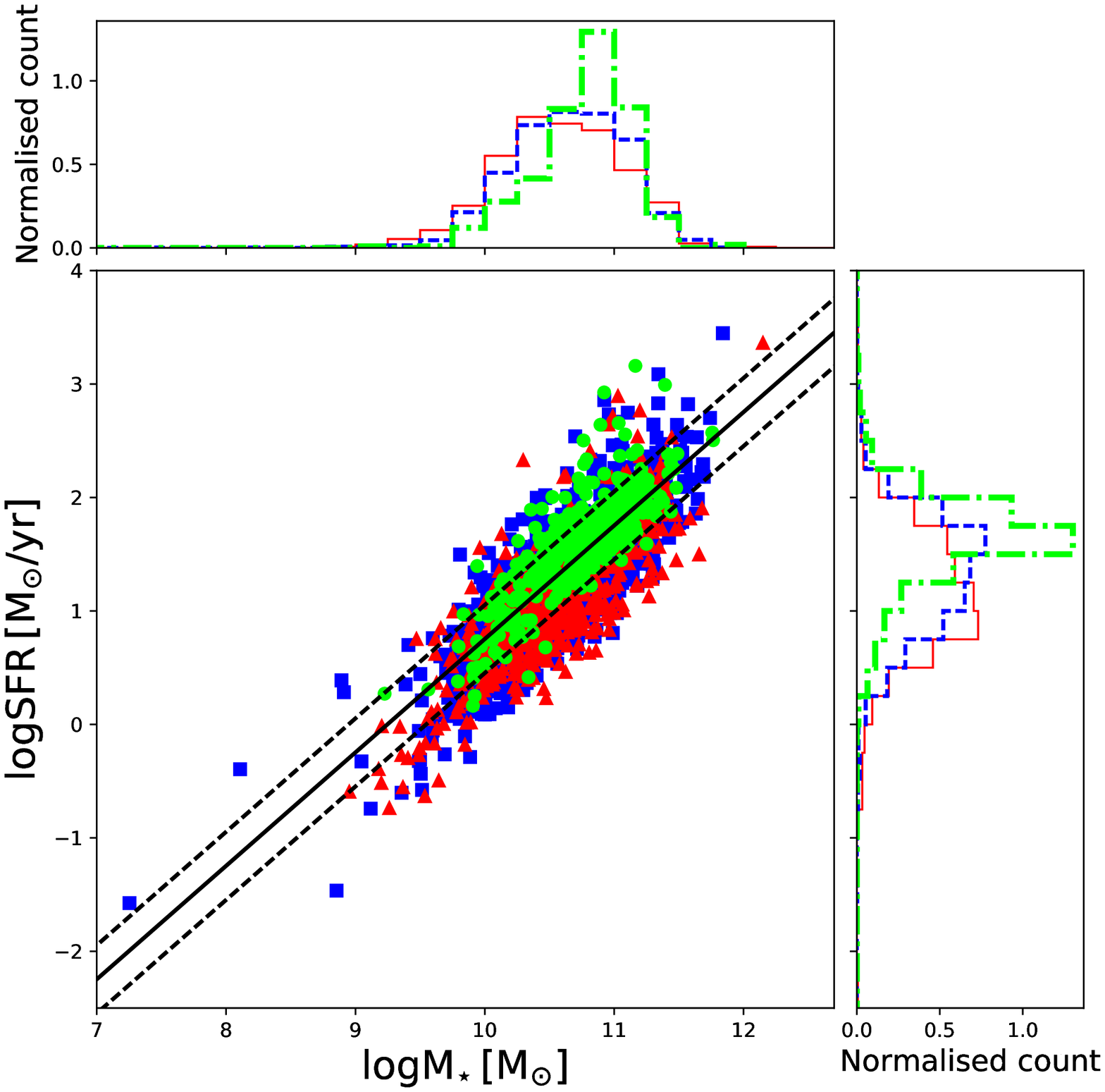}
\end{minipage}
\caption[ ]{Same as Fig.~\ref{MS_both}, representing FIR green valley AGN (left) and non-AGN (right) samples, but in relation with morphology. Top and right histograms of each panel represent the stellar mass and SFR distributions, respectively, of three morphological types.   In all plots, elliptical and lenticular galaxies are represented with red triangles and red solid histograms, spirals with blue squares and blue dashed lines, and irregular and peculiar systems with green filled circles and green dash-dot-dash lines. Black solid and dashed lines show again the main sequence of SF (see Figure ~\ref{MS_both} for all information).
\label{MS_morph}}
\end{figure*}

In Figure ~\ref{MS_morph} we again represent the SFR-stellar mass relation of our AGN (left plot) and non-AGN (right plot) samples, but in relation with morphology. Morphological classification was extracted from \cite{Tasca2009} catalogue. As mentioned in Section ~\ref{Data}, three different classifications are available. In this work we used the one called 'class\_linee' in the catalogue, which is based on 5 morphological parameters including concentration, asymmetry, Gini index, smoothness (or clumpiness), and M20 moment of light (see \citet{Tasca2009} for more information). As it can be seen from the figure we are finding all Hubble types in the green valley. Of all AGN 50, 40, and 13 were classified as E/S0, spiral, and irregular, respectively (602, 1574, and 433 for non-AGN, respectively). Please note that class\_linee classification in \cite{Tasca2009} does not distinguish between irregular galaxies and interactions and mergers. In the next paper we will analyse in more details morphological properties of AGN and non-AGN samples studied in this paper (work in progress). 

\section{DISCUSSION}

\begin{figure}
\centering
\begin{minipage}[c]{0.49\textwidth}  
 \includegraphics[width=7.5cm,angle=0]{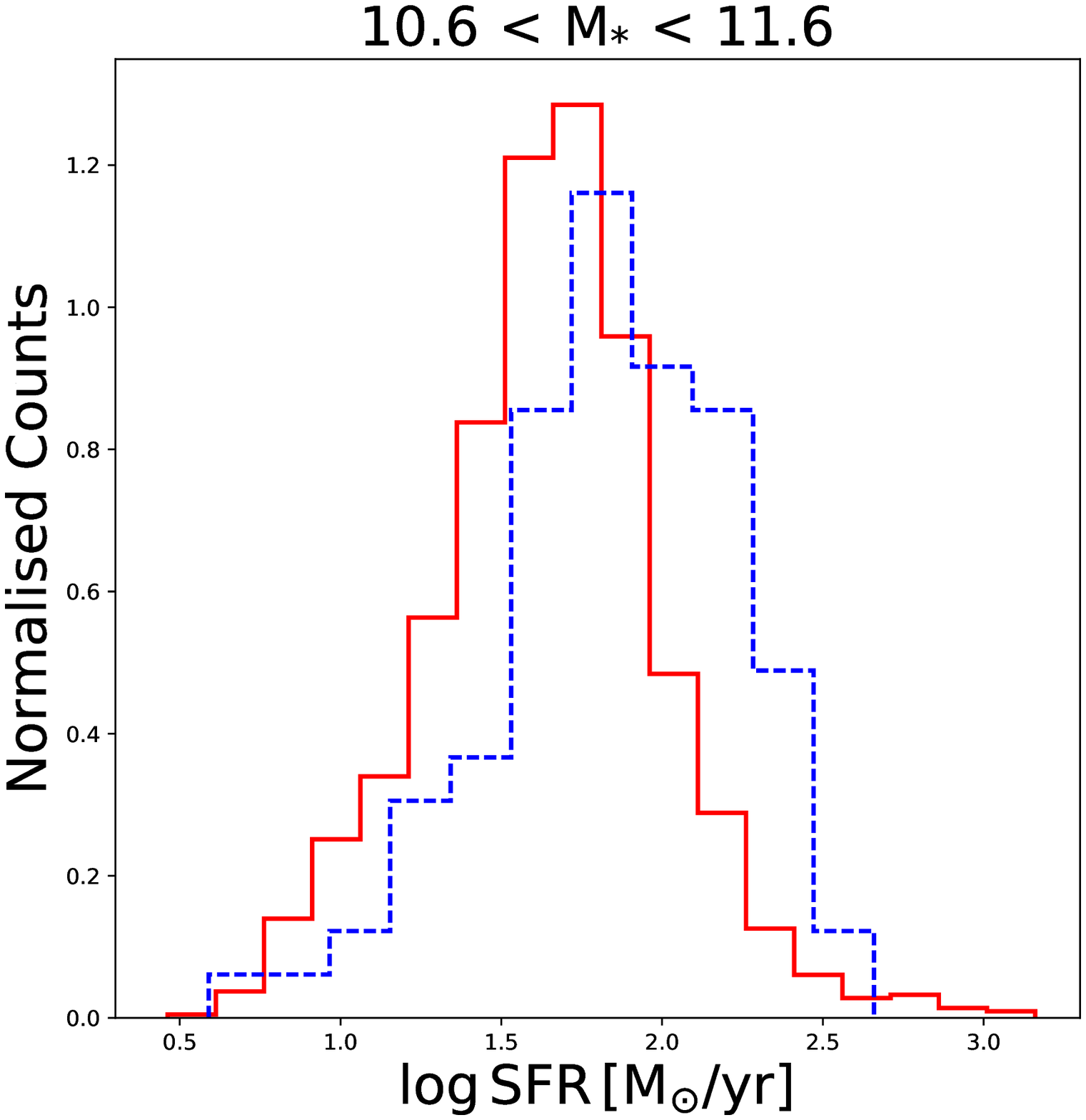}
\end{minipage}
\begin{minipage}[c]{0.49\textwidth}
\includegraphics[width=7.5cm,angle=0]{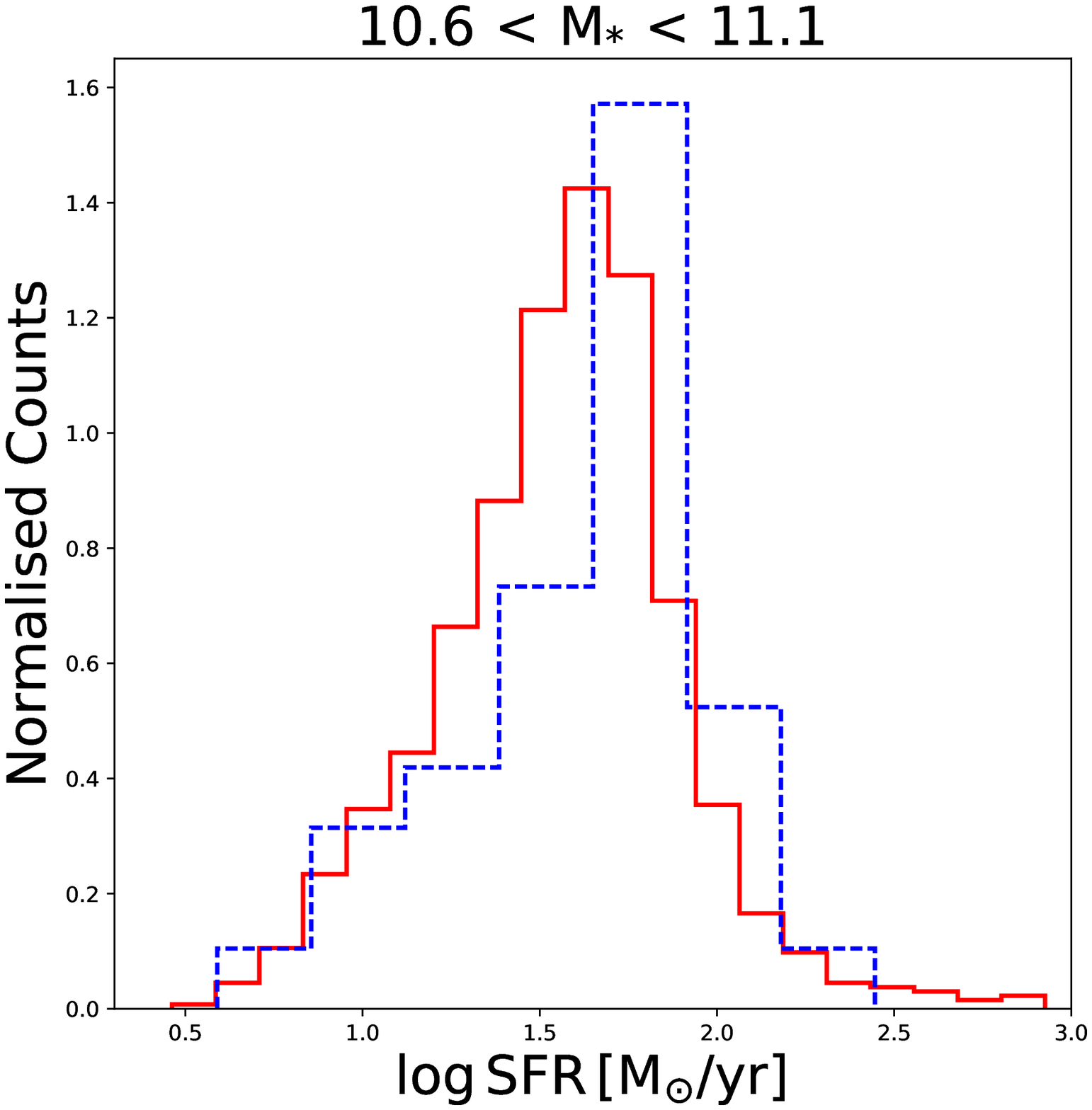}
\end{minipage}
\begin{minipage}[c]{0.49\textwidth}
\includegraphics[width=7.5cm,angle=0]{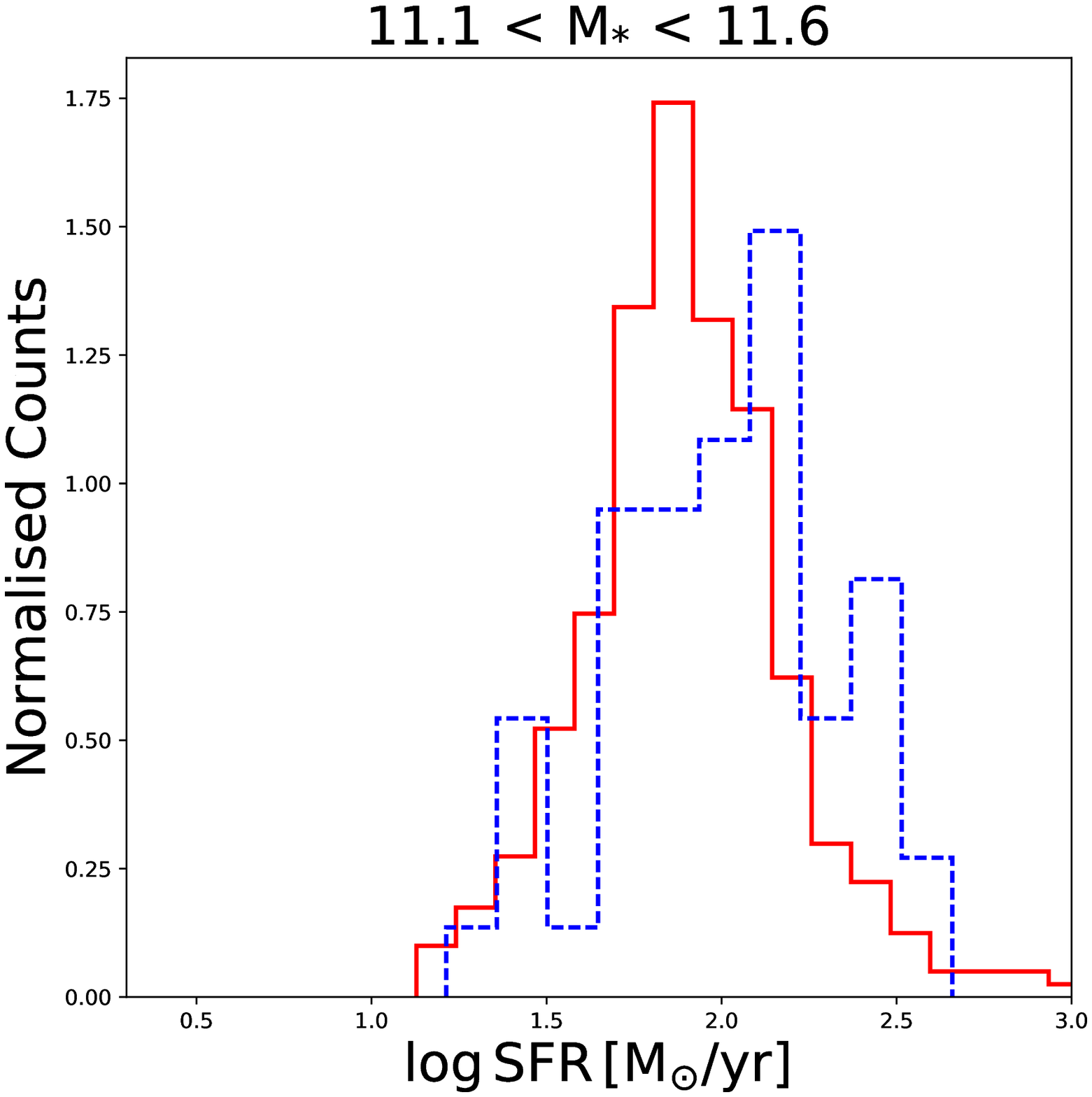}
\end{minipage}
\caption[Normalised distributions of logSFR at fixed mass range]{Normalised distributions of logSFR of AGN (blue dashed histograms) and non-AGN (red solid histograms) for a fixed mass range.}
\label{SameMAssRange}
\end{figure}

\indent Previous works on SF in AGN host galaxies obtained different results when comparing their SF properties with those of inactive galaxies. In some works AGN show elevated SFRs \citep[e.g.,][and references therein]{Rosario},  and in others similar \citep[e.g.,][and references therein]{Stanley2015, Xu2015} or lower \citep[e.g.,][and references therein]{Shimizu2015, Leslie2016} in comparison to non-AGN galaxies. The main reasons for this inconsistencies are related to different data and criteria used to select AGN samples \citep{Hickox2009, Ellison2016} and/or to different methods used to measure their properties \citep[e.g.,][]{kauffmann03, Best2012, Ellison2016}.  \\
\indent As showed above, when comparing the entire samples, we rather observe enhanced SF for AGN galaxies in comparison to non-AGN. However, 
it is known that AGN are hosted by more massive galaxies \citep[e.g.,][]{kauffmann03, Leslie2016, Ellison2016}, which is also the case in our samples \citep[see][]{Nkundabakura}. Therefore, it is crucial to make the comparison of SFRs of AGN and non-AGN samples at fixed stellar mass ranges.
Table~\ref{Table3} shows the median values, first quartile\footnote{Q1, value for which $\sim$\,25\% and $\sim$\,75\% of all sources lie below and above Q1, respectively}, and third quartile\footnote{Q3, value for which $\sim$\,75\% and $\sim$\,25\% of all sources lie below and above Q3, respectively} of SFRs at fixed stellar masses. We analysed the SFRs for the log of stellar masses in the range of 10.6\,-\,11.6 where we have 84\% and 55\% of our FIR green valley AGN and non-AGN galaxies, respectively. Figure~\ref{SameMAssRange} represents the SFR comparisons of AGN and non-AGN at three fixed mass ranges. In all cases we can see that our sample of green valley X-ray detected AGN with FIR emission shows enhanced SF in comparison to non-AGN sample. In particular for masses 10.6\,-\,11.6 (in logarithmic scale) the median values are log(SFR)\,=\,1.85 and log(SFR)\,=\,1.65 for AGN and non-AGN samples, respectively (for values in non-logarithmic scale see Table~\ref{Table3}). These numbers increase for higher stellar masses between 11.1\,-\,11.6 and become 2.01 and 1.88, respectively.
Finally, we performed Kolmogorov-Smirnov statistical test for 10.6\,-\,11.6 mass range and found also that the two distributions are significantly different (with probability parameter of 5.42 $\times \,10^{-6}$).

\begin{table*}
\centering
\caption{Comparison of SFRs between AGN and non-AGN at fixed stellar masses}

\begin{tabular}{|c|c|c|c|c|c|c|c|c|c|}
\hline
\multirow{2}{*}{} & \multicolumn{3}{c|}{$\rm{10.6\leq\,\log (M_{*})\,\leq11.6}$} & \multicolumn{3}{c|}{$\rm{10.6\leq\,\log(M_{*})\,\leq11.1}$} & \multicolumn{3}{c|}{$\rm{11.1\leq\,\log(M_{*})\,\leq11.6}$} \\ \cline{2-10} 
                  & Q1       & Median      & Q3      & Q1       & Median      & Q3      & Q1      & Median     & Q3      \\ \hline
AGN               & 44.66    & 70.79       & 131.82  & 28.84    &47.86        & 77.62   & 58.88   & 102.32     & 151.35        \\ \hline
Non-AGN           & 26.30    & 44.66       & 70.79   & 22.38    & 38.01        & 56.23   & 52.48   & 75.85      & 119.67         \\ \hline
\end{tabular}

\label{Table3}
\end{table*}

\indent Regarding the SF properties of green valley galaxies in general most of works showed that they have SFRs between those of SF and passive galaxies \citep[e.g.,][]{Schiminovich, Salim, Schawinski}. In \citet{Smethurst} the authors found that green valley galaxies are those galaxies which have either left, started to leave the MS of SF, or have still active SF. Moreover, both late- and early-type green valley galaxies show to be located below the MS of SF, at least in the local universe \citep{Schawinski}. To compare AGN and non-AGN samples in relation to the MS of SF, we can use as a reference (at least in the local universe) recent results of \citet{Leslie2016}. The authors classified all SDSS DR7 galaxies (13176 sources in total) into star-forming, composite, Sy2, LINERs, and ambiguous, using the emission line ratios from MPA-JHU DR7 \footnote{http://wwwmpa.mpa-garching.mpg.de/SDSS/DR7/} catalogues. They found that AGN galaxies (Sy2 and LINERs) are mainly located between the MS of SF and passive galaxies (that are located well below the MS). In the case of our sample, as can be seen from both Figure~\ref{MS_both} and Table~\ref{Table1} most of the sources are located on the MS. It is important here to remind that our SFRs where measured using FIR Herschel/PACS data, so we are mapping the brightest sources (both AGN and non-AGN) and also with higher SFRs than in optical \citep{Ellison2016}. However even so, we are observing different trends than in optical when comparing AGN and non-AGN samples, with AGN having higher SFRs, as showed above.  \\
\indent It was proposed that AGN feedback can play a positive or negative role on SF activity \citep[e.g.,][and references therein]{Salom2015, Ciotti2015}: by blowing out the material from the galaxy and quenching (or decreasing) SF, or triggering the SF by compressing dense clouds in the interstellar medium. While the negative AGN feedback is nowadays very well accepted \citep[][and references therein]{Silk2010, Heckman2014}, the positive one was studied only in a scarce number of works \citep[e.g.][and references therein]{Salom2015}. As mentioned above, different works based on X-ray detected AGN samples suggested that AGN negative feedback may play an important role in quenching SF and moving galaxies from the blue cloud to red sequence \citep{Sanchez2004, Nandra, Georgakakis2008, Silverman2008, Treister, Povi2, Povi}. Moreover, taking into account \cite{Leslie2016} optical selection of SDSS galaxies, and the location of AGN and SF galaxies respect to MS, the authors gave the same interpretation that AGN negative feedback is responsible for quenching SF in galaxies. Considering the results obtained in this work where on average the SFRs in AGN are higher in comparison to the same mass non-AGN galaxies we found that: first, green valley X-ray detected AGN that show FIR emission have still very active SF, being located either on or above the MS, and secondly, they do not show signs of SF quenching, but rather its enhancement. Therefore, our results may suggest that for X-ray detected AGN but with FIR emission if there is an influence of AGN feedback on the SF in the green valley, the scenario of AGN positive feedback seems to take place rather than the negative one. \\

\section{CONCLUSIONS}

In this work we present the results of a study of SFRs of green valley X-ray detected AGN and
non-AGN at z\,$<$\,3 (but with $>$\,90\% of sources being at z\,$<$\,1.2) in the COSMOS field. Green valley galaxies were
selected using U\,-\,B rest-frame colour criteria of 0.8\,$\le$\,U\,-\,B\,$\le$\,1.2 with magnitude of $\rm{I_{subaru}\leq}$\,23. We searched for Herschel/PACS FIR data and measured the SFRs of 114 and 2958 green valley AGN and non-AGN galaxies, respectively. We fitted the MIR/FIR SED and measured the total IR luminosity with Le Phare code, by using two different template sets for AGN and non-AGN galaxies. Finally, we obtained good fits for 103 and 2609 AGN and non-AGN, respectively, and measured their SFRs. We compared the SFRs between the two samples and we also observed their location on the main sequence of star formation. Our main results are the following:
\begin{itemize}
\item We observed enhanced SF for AGN galaxies in comparison to non-AGN, when comparing both the entire samples and samples at fixed stellar mass ranges.
\item FIR green valley AGN and non-AGN samples are located on or above the MS of SF, while in most of previous studies based on optical data green valley galaxies were found between the SF and passive galaxies.
\item Regarding the previous point, we found all morphological types present in both samples.

\item AGN studied in this work do not show signs of SF quenching (as suggested in most of previous studies of X-ray detected AGN), but rather its enhancement. Therefore, our results may suggest that for X-ray detected AGN with FIR emission if there is an influence of AGN feedback on the SF in green valley, the scenario of AGN positive feedback seems to rather take the place than the negative one.
\end{itemize}

\section*{Acknowledgements}
We thank the anonimous referee for accepting to review this paper and for giving us constructive comments that improved this work
significantly. We also highly acknowledge the help of Allison Kirpatrick (Yale University) for giving us access to
MIR templates, and of Josefa Masegosa (IAA-CSIC) and Petri Vaisanen (SALT Observatory) for their usefull discussions and comments. Financial support from the Swedish International Development Cooperation Agency (SIDA) through the International Science Programme (ISP)-Uppsala University to University of Rwanda through the Rwanda Astrophysics, Space and Climate Science Research Group (RASCSRG), East African Astrophysics Research Network (EAARN) are gratefully acknowledged. MP acknowledges financial supports from JAE-Doc program of the Spanish National Research Council (CSIC) co-funded by
the European Social Fund, from the Spanish Ministry of Economy and Competitiveness (MINECO) through projects AYA2013-42227-P and AYA2016-76682C3-1-P, and from the Ethiopian Space Science and Technology Institute (ESSTI) under the Ethiopian Ministry of Science and Technology (MoST). We also thank the LePhare team for making their code publicly available.




\bibliographystyle{mnras}
\bibliography{paper_ref} 








\bsp	
\label{lastpage}
\end{document}